\documentclass[preprintnumbers,amsmath,amssymbm,prd]{revtex4}
\usepackage{epsfig}
\usepackage{graphicx}
\usepackage{amssymb}

\begin{document}
\title{Highly excited bound-state resonances of short-range inverse power-law potentials}
%\title{Highly excited bound-state resonances of non-Coulombic inverse power-law potentials}
%\title{Highly excited energy levels of inverse power-law binding potentials}
\author{Shahar Hod}
\affiliation{The Ruppin Academic Center, Emeq Hefer 40250, Israel}
\affiliation{ } \affiliation{The Hadassah Institute, Jerusalem
91010, Israel}
\date{\today}

\begin{abstract}
\ \ \ We study analytically the radial Schr\"odinger equation with
long-range attractive potentials whose asymptotic behaviors are
dominated by inverse power-law tails of the form $V(r)=-\beta_n
r^{-n}$ with $n>2$. In particular, assuming that the effective
radial potential is characterized by a short-range infinitely
repulsive core of radius $R$, we derive a compact {\it analytical}
formula for the threshold energy
$E^{\text{max}}_l=E^{\text{max}}_l(n,\beta_n,R)$ which characterizes
the most weakly bound-state resonance (the most excited energy
level) of the quantum system.
\end{abstract}
\bigskip
\maketitle

%]

\section{Introduction}

The Schr\"odinger differential equation with inverse power-law
attractive potentials has attracted the attention of physicists and
mathematicians since the early days of quantum mechanics. In
particular, long-range power-law potentials play a key role in
theoretical models describing the physical interactions of atoms and
molecules (see \cite{NC1,NC2,NC3,NC4,NC5,NC6,NC7,NC8,LanLif} and
references therein).

It is well known that the attractive Coulombic potential is
characterized by an infinite spectrum $\{E_k\}^{k=\infty}_{k=0}$ of
stationary bound-state resonances with the asymptotic property
$E_{k\to\infty}\to 0^{-}$\cite{LanLif}. On the other hand,
attractive radial potentials whose asymptotic spatial behaviors are
dominated by inverse power-law decaying tails of the form
\begin{equation}\label{Eq1}
V(r)=-{{\beta_n}\over{r^{n}}}\ \ \ \ \text{with}\ \ \ \ n>2
\end{equation}
can only support a {\it finite} number of bound-state resonances
\cite{LanLif}. In particular, it is interesting to note that, for
generic values of the physical parameters $n$ and $\beta_n$, the
discrete energy spectrum of an attractive inverse power-law
potential of the form (\ref{Eq1}) terminates at some finite non-zero
energy $E^{\text{max}}(n,\beta_n)$
\cite{NC1,NC2,NC3,NC4,NC5,NC6,NC7,NC8,LanLif,NoteE0}.

%The characteristic finite energy spectra of attractive power-law
%potentials of the form (\ref{Eq1}) can be computed numerically or
%semi-analytically using a modified WKB method
%\cite{NC1,NC2,NC3,NC4,NC5,NC6,NC7,NC8,LanLif}.
The main goal of the present paper is to present
%an alternative
a simple and elegant mathematical technique for the calculation of
the most excited energy levels $E^{\text{max}}(n,\beta_n)$
\cite{Notewb} which characterize the family (\ref{Eq1}) of
attractive inverse power-law potentials. In particular, below we
shall derive a compact {\it analytical} formula for the threshold
(maximal) energies $E^{\text{max}}(n,\beta_n)$ which characterize
the most weakly bound-state resonances (the most excited energy
levels) of the radial Schr\"odinger equation with the inverse
power-law attractive potentials (\ref{Eq1}) \cite{Notecon}.

\section{Description of the system}

We shall analyze the physical properties of a quantum system whose stationary resonances are determined by the radial Schr\"odinger equation \cite{Notemul}
\begin{equation}\label{Eq2}
\Big[-{{\hbar^2}\over{2\mu}}{{d^2}\over{dr^2}}+{{\hbar^2l(l+1)}\over{2\mu r^2}}+V(r)\Big]\psi_l=E \psi_l\  ,
\end{equation}
where the effective radial potential $V(r)$ in (\ref{Eq2}) is characterized by a long-range inverse power-law attractive part and a short-range infinitely repulsive core. Specifically, we shall consider a composed radial potential of the form
\begin{equation}\label{Eq3}
V(r)=
\begin{cases}
+\infty\ & \ \ \text{for}\ \ \ r\leq R\  ; \\
-{{\beta_n}\over{r^{n}}} & \ \ \text{for}\ \ \ r>R\  .
\end{cases}
\end{equation}

The bound-state ($E<0$) resonances of the Schr\"odinger differential equation (\ref{Eq2}) that we shall analyze in the present paper are characterized by exponentially decaying radial eigenfunctions at spatial infinity:
\begin{equation}\label{Eq4}
\psi_l(r\to\infty)\sim e^{-\kappa r}\  ,
\end{equation}
where \cite{Notew0}
\begin{equation}\label{Eq5}
\kappa^2\equiv -{{2\mu}\over{\hbar^2}}E\ \ \ \ \text{with}\ \ \ \
\kappa\in \mathbb{R}\  .
\end{equation}
In addition, the repulsive core of the effective radial potential (\ref{Eq3}) dictates the inner boundary condition
\begin{equation}\label{Eq6}
\psi_l(r=R)=0\
\end{equation}
for the characteristic radial eigenfunctions.

The Schr\"odinger equation (\ref{Eq2}), supplemented by the radial
boundary conditions (\ref{Eq4}) and (\ref{Eq6}), determine the
discrete spectrum of bound-state eigen-wavenumbers
$\{\kappa(n,\beta_n,R)\}$ [or equivalently, the discrete spectrum of
binding energies $E(n,\beta_n,R)$] which characterize the effective
radial potential (\ref{Eq3}). As we shall explicitly show in the
next section, the most weakly bound-state resonance (that is, the
most excited energy level) which characterizes the quantum system
(\ref{Eq3}) can be determined {\it analytically} in the regime
\cite{Notese,Notel0n,Grib}
\begin{equation}\label{Eq7}
\kappa r_n\ll1\
\end{equation}
of small binding energies, where the characteristic length-scale
$r_n$ is defined by the relation
\begin{equation}\label{Eq8}
r_n\equiv \Big[{{2\mu\beta_n}\over{(n-2)^2\hbar^2}}\Big]^{1/(n-2)}\
.
\end{equation}

\section{The resonance equation and its regime of validity}

In the present section we shall analyze the radial Schr\"odinger equation
\begin{equation}\label{Eq9}
\Big\{{{d^2}\over{dr^2}}-{{1}\over{r^2}}\Big[(\kappa
r)^2+l(l+1)-(n-2)^2\Big({{r_n}\over{r}}\Big)^{n-2}\Big]\Big\}\psi_l(r;\kappa,r_n,n)=0\
,
\end{equation}
which determines the spatial behavior of the bound-state
eigenfunctions $\psi_l(r)$ in the regime $r>R$. As we shall
explicitly show below, the characteristic radial equation
(\ref{Eq9}) can be solved analytically in the two asymptotic radial
regions $r\ll 1/\kappa$ and $r\gg r_n$. We shall then show that, for
small resonant energies in the regime $\kappa r_n\ll1$ [see
(\ref{Eq7}) \cite{Notese}], one can use a functional matching
procedure in the overlapping region $r_n\ll r\ll 1/\kappa$ in order
to determine the binding energies $\{E(n,r_n,R,l)\}$ [or
equivalently, the eigen-wavenumbers $\{\kappa(n,r_n,R,l)\}$] which
characterize the marginally bound-state resonances of the radial
Schr\"odinger equation (\ref{Eq2}) with the effective binding
potential (\ref{Eq3}).

We shall first solve the Schr\"odinger equation (\ref{Eq9}) in the radial region
\begin{equation}\label{Eq10}
r\ll 1/\kappa\  ,
\end{equation}
in which case one may approximate (\ref{Eq9}) by
\begin{equation}\label{Eq11}
\Big[{{d^2}\over{dr^2}}-{{l(l+1)}\over{r^2}}+{{(n-2)^2r^{n-2}_n}\over{r^n}}\Big]\psi_l=0\
.
\end{equation}
The general solution of the radial differential equation (\ref{Eq11}) can be expressed in terms of the Bessel
functions of the first and second kinds (see Eq. 9.1.53 of \cite{Abram}):
\begin{equation}\label{Eq12}
\psi_l(r)=A_1r^{1\over
2}J_{{2l+1}\over{n-2}}\Big[2\Big({{r_n}\over{r}}\Big)^{(n-2)/2}\Big]+A_2r^{1\over
2}Y_{{2l+1}\over{n-2}}\Big[2\Big({{r_n}\over{r}}\Big)^{(n-2)/2}\Big]\
,
\end{equation}
where $\{A_1,A_2\}$ are normalization constants to be determined
below. Using the small-argument ($r_n/r\ll1$) asymptotic behaviors
of the Bessel functions (see Eqs. 9.1.7 and 9.1.9 of \cite{Abram}),
one finds from (\ref{Eq12}) the expression
\begin{eqnarray}\label{Eq13}
\psi_l(r)=A_1
{{r^{{1}/{2}}_n}\over{\big({{2l+1}\over{n-2}}\big)\Gamma\big({{2l+1}\over{n-2}}\big)}}\Big({{r_n}\over{r}}\Big)^l-A_2
{{r^{{1}/{2}}_n\Gamma\big({{2l+1}\over{n-2}}\big)}\over{\pi}}\Big({{r}\over{r_n}}\Big)^{l+1}\
\end{eqnarray}
for the radial eigenfunction which characterizes the weakly-bound (highly-excited) states of the
Schr\"odinger differential equation (\ref{Eq9}) in the intermediate radial region
\begin{equation}\label{Eq14}
r_n\ll r \ll 1/\kappa\  .
\end{equation}

We shall next solve the Schr\"odinger equation (\ref{Eq9}) in the radial region
\begin{equation}\label{Eq15}
r\gg r_n\  ,
\end{equation}
in which case one may approximate (\ref{Eq9}) by
\begin{equation}\label{Eq16}
\Big[{{d^2}\over{dr^2}}-\kappa^2-{{{l(l+1)}}\over{r^2}}\Big]\psi_l=0\
.
\end{equation}
The general solution of the radial differential equation (\ref{Eq16}) can be expressed in terms of the Bessel
functions of the first and second kinds (see Eq. 9.1.49 of \cite{Abram}):
\begin{equation}\label{Eq17}
\psi_l(r)=B_1r^{1\over 2}J_{l+{{1}\over{2}}}(i\kappa r)+B_2r^{1\over
2}Y_{l+{{1}\over{2}}}(i\kappa r)\ ,
\end{equation}
where $\{B_1,B_2\}$ are normalization constants \cite{Notemp}. Using
the small-argument ($\kappa r\ll1$) asymptotic behaviors of the
modified Bessel functions (see Eqs. 9.1.7 and 9.1.9 of
\cite{Abram}), one finds from (\ref{Eq17}) the expression
\begin{eqnarray}\label{Eq18}
\psi_l(r)=B_1{{({{i\kappa}/{2}})^{l+{{1}\over{2}}}}\over{\big(l+{{1}\over{2}}\big)\Gamma\big(l+{{1}\over{2}}\big)}}
r^{l+1}-B_2
{{\Gamma\big(l+{{1}\over{2}}\big)}\over{\pi({{i\kappa}/{2}})^{l+{{1}\over{2}}}}}r^{-l}\
\end{eqnarray}
for the radial eigenfunction which characterizes the weakly bound-state resonances (the highly-excited states) of the
Schr\"odinger differential equation (\ref{Eq9}) in the intermediate radial region
\begin{equation}\label{Eq19}
r_n\ll r \ll 1/\kappa\  .
\end{equation}

Interestingly, for weakly bound-state resonances (that is, for small
resonant wave-numbers in the regime $\kappa r_n\ll1$),  the two
expressions (\ref{Eq13}) and (\ref{Eq18}) for the characteristic
eigenfunction $\psi_l(r)$ of the radial Schr\"odinger equation
(\ref{Eq9}) are both valid in the intermediate radial region $r_n\ll
r \ll 1/\kappa$ [see Eqs. (\ref{Eq14}) and (\ref{Eq19})]. Note, in
particular, that these two analytical expressions for the radial
eigenfunction $\psi_l(r)$ are characterized by the same functional
(radial) behavior. One can therefore express the coefficients
$\{B_1,B_2\}$ of the radial solution (\ref{Eq17}) in terms of the
coefficients $\{A_1,A_2\}$ of the radial solution (\ref{Eq12}) by
matching the two mathematical expressions (\ref{Eq13}) and
(\ref{Eq18}) for the characteristic radial eigenfunction $\psi_l(r)$
in the intermediate radial region $r_n\ll r \ll 1/\kappa$. This
functional matching procedure yields the relations \cite{Notesr}
\begin{equation}\label{Eq20}
B_1=-A_2
{{\big(l+{{1}\over{2}}\big)\Gamma\big(l+{{1}\over{2}}\big)\Gamma
\big({{2l+1}\over{n-2}}\big)}\over{\pi}}\Big({{2}\over{i\kappa
r_n}}\Big)^{l+{{1}\over{2}}}\
\end{equation}
and
\begin{equation}\label{Eq21}
B_2=-A_1
{{\pi}\over{\big({{2l+1}\over{n-2}}\big)\Gamma\big(l+{{1}\over{2}}\big)\Gamma\big({{2l+1}\over{n-2}}\big)}}\Big({{i\kappa
r_n}\over{2}}\Big)^{l+{{1}\over{2}}}\  .
\end{equation}

We are now in a position to derive the resonance equation which
determines the binding energies $\{E(n,r_n,l)\}$ [or equivalently,
the eigen-wavenumbers $\{\kappa(n,r_n,l)\}$] of the weakly-bound
(highly-excited) states which characterize the radial Schr\"odinger
equation (\ref{Eq2}) with the effective radial potential
(\ref{Eq3}). Using Eqs. 9.2.1 and 9.2.2 of \cite{Abram}, one finds
the asymptotic spatial behavior
\begin{equation}\label{Eq22}
\psi_l(r\to\infty)=B_1\sqrt{2/i\pi\kappa}\cdot\cos(i\kappa
r-l\pi/2-\pi/2)+B_2\sqrt{2/i\pi\kappa}\cdot\sin(i\kappa
r-l\pi/2-\pi/2)\
\end{equation}
for the radial eigenfunction (\ref{Eq17}). Taking cognizance of the boundary condition (\ref{Eq4}), which
characterizes the bound-state resonances of the radial Schr\"odinger equation (\ref{Eq2}), one deduces from
(\ref{Eq22}) the simple relation \cite{Noterc}
\begin{equation}\label{Eq23}
B_2=iB_1\  .
\end{equation}
Substituting Eqs. (\ref{Eq20}) and (\ref{Eq21}) into (\ref{Eq23}), one obtains the characteristic resonance equation
%\cite{Notega}
\begin{equation}\label{Eq24}
\Big({{i\kappa
r_n}\over{2}}\Big)^{2l+1}=i{{2}\over{n-2}}\Bigg[{{\big(l+{{1}\over{2}}\big)\Gamma\big(l+{{1}\over{2}}\big)\Gamma\big({{2l+1}\over{n-2}}\big)}\over{\pi}}\Bigg]^2
\cdot{{A_2}\over{A_1}}
\end{equation}
for the highly-excited bound-state resonances
which characterize the Schr\"odinger equation (\ref{Eq2}) with the effective radial potential (\ref{Eq3}).

\section{The resonant binding energy of the most excited energy level}

The dimensionless ratio $A_2/A_1$ that appears in the resonance equation (\ref{Eq24}) can be determined by the inner boundary condition (\ref{Eq6}) which is dictated by the short-range repulsive part of the effective radial potential (\ref{Eq3}). In particular, substituting (\ref{Eq12}) into (\ref{Eq6}), one finds
\begin{equation}\label{Eq25}
{{A_2}\over{A_1}}=-{{J_{{2l+1}\over{n-2}}\Big[2\big({{r_n}\over{R}}\big)^{(n-2)/2}\Big]}
\over{Y_{{2l+1}\over{n-2}}\Big[2\big({{r_n}\over{R}}\big)^{(n-2)/2}\Big]}}\
.
\end{equation}
Substituting the dimensionless ratio (\ref{Eq25}) into the resonance
equation (\ref{Eq24}), one finally finds the expression
\cite{Noteint}
%\cite{Noteps}
\begin{equation}\label{Eq26}
\kappa
r_n=\Bigg\{{{(-1)^{l+1}}\over{n-2}}\Bigg[{{2^{l+1}\big(l+{{1}\over{2}}\big)\Gamma\big(l+{{1}
\over{2}}\big)\Gamma\big({{2l+1}\over{n-2}}\big)}\over{\pi}}\Bigg]^2
\cdot{{J_{{2l+1}\over{n-2}}\Big[2\big({{r_n}\over{R}}\big)^{(n-2)/2}\Big]}
\over{Y_{{2l+1}\over{n-2}}\Big[2\big({{r_n}\over{R}}\big)^{(n-2)/2}\Big]}}\Bigg\}^{1/(2l+1)}\
\end{equation}
for the dimensionless resonant wave-number which characterizes the
most excited energy level (the most weakly bound-state resonance) of
the radial Schr\"odinger equation (\ref{Eq2}) with the effective
binding potential (\ref{Eq3}).

It is worth emphasizing again that the analytically derived resonance equation (\ref{Eq24})
is valid in the regime [see (\ref{Eq14}) and (\ref{Eq19})] \cite{Noterov}
\begin{equation}\label{Eq27}
\kappa r_n\ll1\
\end{equation}
of small binding energies. Taking cognizance of Eq. (\ref{Eq26}),
one realizes that the small wave-number requirement (\ref{Eq27}) is
satisfied for
%in the two physically interesting regimes:
%\newline
%(1) In the regime
%\begin{equation}\label{Eq30}
%{{2}\over{n-2}}\Big({{r_n}\over{R}}\Big)^{(n-2)/2}\ll1\  ,
%\end{equation}
%in which case one finds [see Eq. (\ref{Eq}) and Eqs. 9.1.7 and 9.1.9 of \cite{Abram}]
%\begin{equation}\label{Eq29}
%\kappa R=\Big\{{{(-1)^{l}}\over{\pi(2l+1)}}\Big[2^{l+1}\Gamma\big(l+{{3}\over{2}}\big)\Big]^2\Big\}^{1/(2l+1)}\
%\end{equation}
\begin{equation}\label{Eq28}
2\Big({{r_n}\over{R}}\Big)^{(n-2)/2}\simeq j_{{{2l+1}\over{n-2}},k}\
\ ,
\end{equation}
where $\{j_{\nu,k}\}^{k=\infty}_{k=1}$ are the positive zeros of the
Bessel function $J_{\nu}(x)$ \cite{Abram,Bes}. Defining the
dimensionless small quantity
\begin{equation}\label{Eq29}
\Delta_k\equiv2\Big({{r_n}\over{R}}\Big)^{(n-2)/2}-j_{{{2l+1}\over{n-2}},k}\ll1\
,
\end{equation}
one finds from (\ref{Eq26}) the expression \cite{Notete,Notesg}
\begin{equation}\label{Eq30}
\kappa
r_n=\Bigg\{{{(-1)^{l}}\over{\pi(n-2)}}\Big[{{(2l+1)!!\Gamma\big({{2l+1}\over{n-2}}\big)}}\Big]^2
\cdot{{J_{{{2l+1}\over{n-2}}+1}\big(j_{{{2l+1}\over{n-2}},k}\big)}\over{Y_{{2l+1}\over{n-2}}\big(j_{{{2l+1}\over{n-2}},k}\big)}}\Delta_k\Bigg\}^{1/(2l+1)}\
\end{equation}
for the smallest resonant wave-number which characterizes the
effective binding potential (\ref{Eq3}).

\section{Summary}

We have studied analytically the Schr\"odinger differential equation
with attractive radial potentials whose asymptotic behaviors are
dominated by inverse power-law tails of the form $V(r)=-\beta_n
r^{-n}$ with $n>2$. These long-range radial potentials are of great
importance in physics and chemistry. In particular, they provide a
quantitative description for the physical interactions of atoms and
molecules \cite{NC1,NC2,NC3,NC4,NC5,NC6,NC7,NC8,LanLif}.

Using a low-energy matching procedure, we have derived the analytical expression [see Eqs. (\ref{Eq5}), (\ref{Eq8}), (\ref{Eq29}), and (\ref{Eq30})]
\begin{equation}\label{Eq31}
\Big({{2\mu\beta^{{{2}\over{n}}}_n}\over{\hbar^2}}\Big)^{{{n}\over{n-2}}}\cdot E^{\text{max}}(n,l)=-\Bigg\{{{(-1)^{l}}\over{\pi(n-2)}}\Big[{{(2l+1)!!(n-2)^{{2l+1}\over{n-2}}\Gamma\big({{2l+1}\over{n-2}}\big)}}\Big]^2 \cdot{{J_{{{2l+1}\over{n-2}}+1}\big(j_{{{2l+1}\over{n-2}},k}\big)}\over{Y_{{2l+1}\over{n-2}}\big(j_{{{2l+1}\over{n-2}},k}\big)}}\Delta_k\Bigg\}^{{{2}\over{2l+1}}}\
\end{equation}
for the dimensionless threshold energy which characterizes the most
excited energy level (the most weakly bound-state resonance) of the
radial Schr\"odinger equation (\ref{Eq2}) with the effective binding
potential (\ref{Eq3}). It is worth noting that in the regime $n\gg1$
\cite{Notenum} of fast decaying inverse power-law potentials, one
finds from (\ref{Eq31}) the compact formula \cite{Notegz,Notel0}
\begin{equation}\label{Eq32}
{{2\mu\beta^{{{2}\over{n}}}_n}\over{\hbar^2}}\cdot E^{\text{max}}(n\gg1,l)=
-\Big\{{{(-1)^{l}[(2l-1)!!]^2n}\over{\pi}} \cdot{{J_1(j_{0,k})}\over{Y_0(j_{0,k})}}\Delta_k\Big\}^{{{2}\over{2l+1}}}\
\end{equation}
for the characteristic threshold energy of the most excited
bound-state resonance.

As a consistency check, it is worth mentioning that, in the special
case of spherically symmetric ($l=0$) wave functions, Eq.
(\ref{Eq24}) reduces to the semi-classical result
$\kappa=1/(a-\bar{a})$ of \cite{Grib}, where $a$ is the s-wave
scattering length and $\bar{a}=\pi
r_n(n-2)\cot[\pi/(n-2)]/\Gamma^2[1/(n-2)]$ \cite{Notetf}. In
addition, it is worth emphasizing that the interesting work
presented in \cite{Grib} for the $l=0$ case is based on the
semi-classical WKB analysis, whereas in the present paper we have
presented a full quantum-mechanical treatment of the physical system
which is valid for generic values of the dimensionless angular
momentum parameter $l$.

\bigskip
\noindent
{\bf ACKNOWLEDGMENTS}
\bigskip

This research is supported by the Carmel Science Foundation. I thank
Oded Hod for helpful discussions. I would also like to thank Yael
Oren, Arbel M. Ongo, Ayelet B. Lata, and Alona B. Tea for
stimulating discussions.

%\newpage


\begin{thebibliography}{99}
\bibitem{Notewb} Note that the most excited energy levels $E^{\text{max}}(n,\beta_n)$ correspond to the most weakly
bound-state resonances of the inverse power-law binding potentials (\ref{Eq1}).
\bibitem{Notecon} It is worth mentioning that the weakly bound-state resonances which characterize the inverse power-law potentials (\ref{Eq1}) play a key role in the physical description of the scattering of low-energy atoms and molecules and in widely used theoretical models of the Bose-Einstein condensation phenomenon \cite{NC8}.
\bibitem{Notemul} Here $\mu$ is the effective (reduced) mass of the physical system and $l$ is the spherical harmonic index of the quantum mode.
\bibitem{Notew0} We shall assume, without loss of generality, that $\kappa>0$.
\bibitem{Notese} Note that the strong inequality (\ref{Eq7}) corresponds to small resonant energies in the regime
$(2\mu\beta^{2/n}_n/\hbar^2)^{n/(n-2)}E\ll1$ [see Eqs. (\ref{Eq5}) and (\ref{Eq8})].
\bibitem{Notemp} As we shall explicitly show below, the normalization constants $\{B_1,B_2\}$ of the
radial solution (\ref{Eq17}) can be determined by a functional matching procedure.
\bibitem{Notesr} Note that the coefficients $\{A_1,A_2\}$ and $\{B_1,B_2\}$ of the radial solutions (\ref{Eq12}) and (\ref{Eq17})
satisfy the compact relation $B_1B_2/A_1A_2=(n-2)/2$.
\bibitem{Noterov} Since each inequality in (\ref{Eq14}) and (\ref{Eq19}) roughly corresponds to an order-of-magnitude
difference between two physical quantities [that is, $r_n/r\lesssim
10^{-1}$ and $r/(1/\kappa)\lesssim 10^{-1}$ in (\ref{Eq14}) and
(\ref{Eq19})], the analytically derived resonance condition
(\ref{Eq24}) for the weakly bound-state resonances (the most excited
energy levels) of the effective radial potential (\ref{Eq3}) is
expected to be valid in the low wave-number regime $\kappa
r_n\lesssim 10^{-2}$.
\bibitem{Notete} Here we have used the Taylor expansions $J_{{2l+1}\over{n-2}}\big[{{2}\over{n-2}}\big({{r_n}\over{R}}\big)^{(n-2)/2}\big]=J_{{2l+1}\over{n-2}}\big(j_{{{2l+1}\over{n-2}},k}+\Delta_k\big)=J_{{2l+1}\over{n-2}}\big(j_{{{2l+1}\over{n-2}},k}\big)+J^{'}_{{2l+1}\over{n-2}}\big(j_{{{2l+1}\over{n-2}},k}\big)
\cdot\Delta_k+O(\Delta^2_k)=-J_{{{2l+1}\over{n-2}}+1}\big(j_{{{2l+1}\over{n-2}},k}\big)
\cdot\Delta_k+O(\Delta^2_k)$ [see Eq. 9.1.27d of \cite{Abram}] and $Y_{{2l+1}\over{n-2}}\big[{{2}\over{n-2}}\big({{r_n}\over{R}}\big)^{(n-2)/2}\big]=Y_{{2l+1}\over{n-2}}\big(j_{{{2l+1}\over{n-2}},k}+\Delta_k\big)=Y_{{2l+1}\over{n-2}}\big(j_{{{2l+1}\over{n-2}},k}\big)+O(\Delta_k)$. In addition, we have used here Eq. 6.1.12 of \cite{Abram}.
\bibitem{Notesg} It can be checked directly that the
dimensionless ratio ${{J_{{{2l+1}\over{n-2}}+1}(j_{{{2l+1}\over{n-2}},k})}/{Y_{{2l+1}\over{n-2}}(j_{{{2l+1}\over{n-2}},k})}}$ in (\ref{Eq30}) is a positive definite expression. Thus, the assumption $\kappa>0$ (see \cite{Notew0}) corresponds to the relation $(-1)^l\cdot\Delta_k>0$.
\bibitem{Notenum} Note that this asymptotic limit corresponds to the regime
in which a direct numerical integration
of the Schr\"odinger radial equation (\ref{Eq2}) with the effective binding
potential (\ref{Eq3}) becomes a nontrivial exercise.
\bibitem{Notel0} It is worth noting that this expression can be further
simplified in the spherically symmetric ($l=0$) case, in which case one finds from (\ref{Eq32})
the compact expression
${{2\mu\beta^{{{2}/{n}}}_n}\over{\hbar^2}}\cdot E^{\text{max}}(n\gg1,l=0)=-\big[{{n}\over{\pi}}
\cdot{{J_1(j_{0,k})}\over{Y_0(j_{0,k})}}\Delta_k\big]^{2}$ for the characteristic
threshold energy of the most excited quantum level.
\bibitem{Notetf} I would like to thank the anonymous referee for pointing
out with explicit calculations this physically important fact. To
see this relation, one can take $A_1=(1-{\bar
a}/a)(n-2)^{-1+1/(n-2)}\Gamma[1/(n-2)]$ and
$A_2=\pi/\{a(n-2)^{1/(n-2)}\Gamma[1/(n-2)]\}$ in Eq. (\ref{Eq12}),
where ${\bar
a}=\pi\cot[\pi/(n-2)]/\{(n-2)^{-1+2/(n-2)}\Gamma^2[1/(n-2)]\}$. This
would yield the zero-energy $l=0$ asymptotic $r\to\infty$ solution
$\psi_0(r)=1-r/a$, where $a$ is the s-wave scattering length.
Substituting the dimensionless ratio $A_2/A_1=\pi/\{(a-{\bar
a})(n-2)^{-1+2/(n-2)}\Gamma^2[1/(n-2)]\}$ into Eq. (\ref{Eq24}), one
finds $\kappa=1/(a-{\bar a})$ for the spherically symmetric $l=0$
modes, in agreement with \cite{Grib}.
\end{thebibliography}
\end{document}